\documentclass[aps,prx,twocolumn,superscriptaddress]{revtex4-2}
\usepackage{amsmath}
\usepackage{amssymb} 
\usepackage{amsthm}
\usepackage{graphicx}
\usepackage{dcolumn}
\usepackage{bm}
\usepackage{braket}
\usepackage{xcolor}
\usepackage[normalem]{ulem} 
\usepackage{hyperref}
\usepackage[capitalise]{cleveref}
\newcommand{\abs}[1]{\left\vert#1\right\vert}
\newcommand{\norm}[1]{\left\Vert#1\right\Vert}
\newcommand{\comm}[1]{\left[#1\right]}
\newtheorem{theorem}{Theorem}
\newtheorem{lemma}[theorem]{Lemma}

\newcommand{\subI}{_{\mathrm{I}}}
\newcommand{\subM}{_{\mathrm{M}}}
\newcommand{\subF}{_{\mathrm{F}}}
\newcommand{\lambdals}{\lambda_{l,\sigma}}

\DeclareMathOperator*{\Circ}{\bigcirc}

\begin{document}

\title{Lightcone shading for classically accelerated quantum error mitigation}

\author{Andrew Eddins}
\email{aeddins@ibm.com}
\affiliation{IBM Quantum, IBM Research Cambridge, Cambridge, MA 02139, USA}

\author{Minh C. Tran}
\affiliation{IBM Quantum, IBM Research Cambridge, Cambridge, MA 02139, USA}

\author{Patrick Rall}
\affiliation{IBM Quantum, IBM Research Cambridge, Cambridge, MA 02139, USA}

\date{\today}

\begin{abstract}
Quantum error mitigation (QEM) can recover accurate expectation values from a noisy quantum computer by trading off bias for variance, such that an averaged result is more accurate but takes longer to converge.
Probabilistic error cancellation (PEC) stands out among QEM methods as an especially robust means of controllably eliminating bias.
However, PEC often exhibits a much larger variance than other methods, inhibiting application to large problems for a given error rate.
Recent analyses have shown that the variance of PEC can be reduced by not mitigating errors lying outside the causal lightcone of the desired observable \cite{tran2023locality}.
Here, we improve the lightcone approach by classically computing tighter bounds on how much each error channel in the circuit can bias the final result.
This set of bounds, which we refer to as a ``shaded lightcone,’’ enables a more targeted application of PEC, improving the tradespace of bias and variance, while illuminating how the structure of a circuit determines the difficulty of error-mitigated computation.
Although a tight shaded lightcone is exponentially hard to compute, we present an algorithm providing a practical benefit for some problems even with modest classical resources, leveraging the ease of evolving an error instead of the state or the observable.
The algorithm reduces the runtime that would be needed to apply PEC for a target accuracy in an example 127-qubit Trotter circuit by approximately two orders of magnitude compared to standard lightcone-PEC, expanding the domain of problems that can be computed via direct application of PEC on noisy hardware.

\end{abstract}

\maketitle

\section{Introduction}

As quantum processors become capable of estimating expectation values of large numbers of entangled qubits~\cite{Kim2023}, quantum and classical results can be meaningfully benchmarked against one another in terms of accuracy and speed \cite{anand2023classical}. The costs of classical and error-mitigated quantum approaches both grow exponentially with problem size, so the prospect of quantum advantage without error correction may hinge on the arguments of these exponentials \cite{Takagi2022-bd, Kechedzhi2023}.
Inversely, the nominal cost of error mitigation decays exponentially to zero as the hardware error rate improves. Thus for a sufficiently low error rate, this cost can in principle be smaller than the classical counterpart. 

Besides the error rate, the runtime cost, or sampling cost, of mitigation is also sensitive to how efficiently the particular error mitigation method transforms bias into variance, such that progress on this front stands to greatly expand near-term quantum capabilities. 
Zero Noise Extrapolation (ZNE)~\cite{PhysRevLett.119.180509, PhysRevX.7.021050} is a leading error mitigation method with relatively low sampling cost, that typically works by assuming the expectation value varies with error rate as a simple extrapolating function, such as an exponential decay.
However, such an assumption is not guaranteed and can fail even in simple cases~\cite{caiMultiexponentialErrorExtrapolation2021}. Given a local and learnable noise model, Probabilistic Error Cancellation (PEC)---another leading method---precisely injects ``antinoise'' \cite{niroula2023thresholds} throughout the circuit such that, on average, hardware errors are exactly cancelled where they occur in the circuit. By cancelling errors at the source, no simplifying assumptions need be made about how the errors impact the expectation value, and the complete elimination of the bias is mathematically guaranteed~\cite{PhysRevLett.119.180509, Van_den_Berg2023-vw} in the limit of perfect noise characterization \cite{govia2024bounding}.
 However, the rigorous performance guarantee of PEC comes at a high cost, as mitigating the effect of each error on the quantum state requires more resources than mitigating the combined effect of all errors on a single expectation value, as with e.g. ZNE \cite{filippov2024scalability}.
 As more antinoise is added to cancel every error channel individually, the statistical variance of PEC balloons with a particularly severe exponential, limiting applicability to small problems compared to ZNE.

\begin{figure}[t]
    \includegraphics[width=0.85\linewidth]{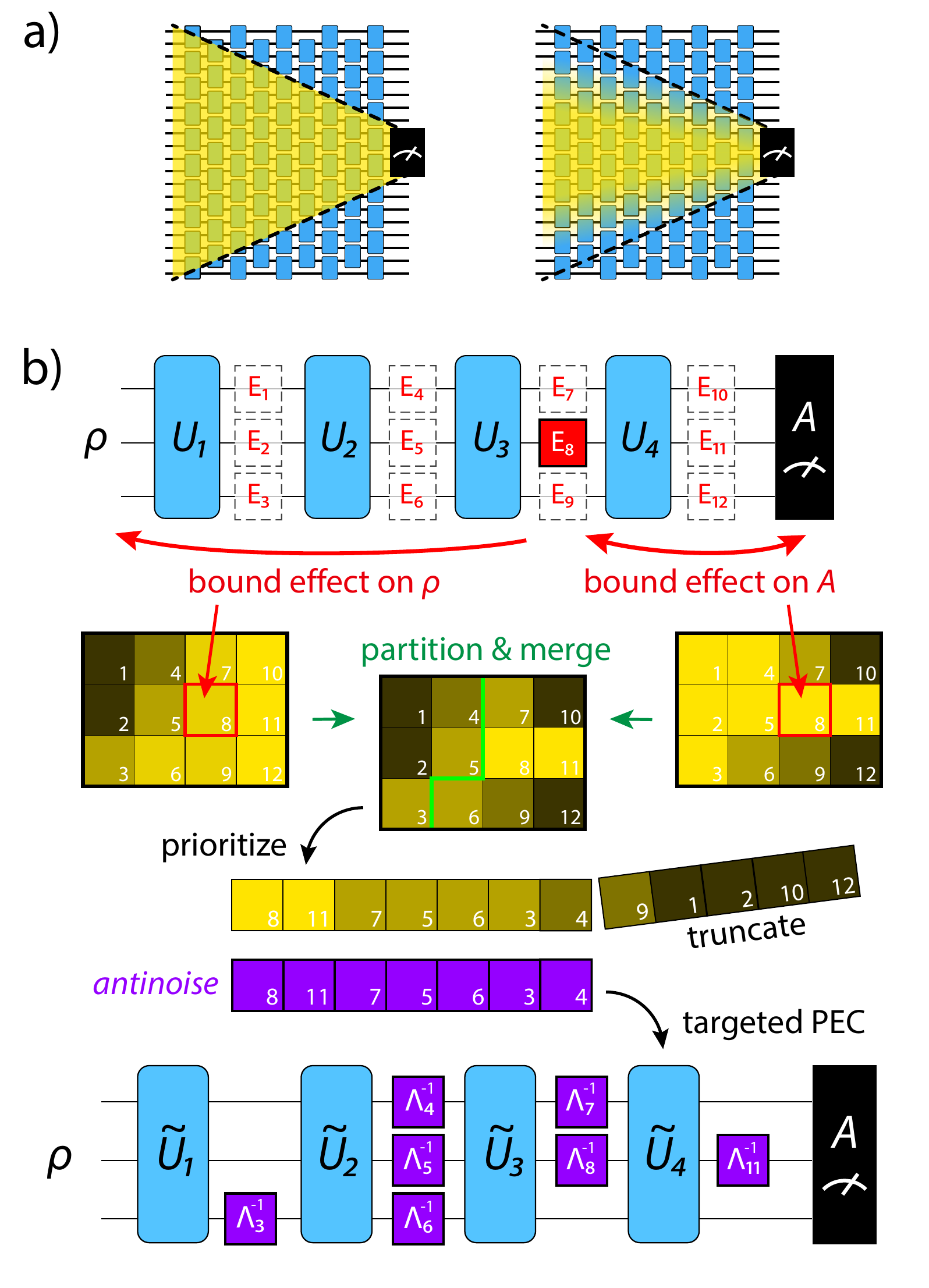}
    \caption{a) A conventional lightcone assigns a binary value to each error channel in a noisy quantum circuit, indicating
    which errors (yellow region)
    can possibly influence a measurement given the topology of gates in the circuit and their commutation relations. A ``shaded lightcone'' generalizes this notion by assigning continuous values that more tightly upper bound the bias from each error channel. The tighter bounds permit more efficiently targeted application of PEC. b) Overview of our algorithm with a fictitious example. We classically bound how each error $E_i$ in our model of the noisy circuit can change the expectation value, first by evolving errors where possible, then by using quantum speed limits to propagate information about $A$ further backwards. We sort the errors by a priority, then mitigate only high-priority errors on the noisy hardware.}
    \label{fig_slc}
\end{figure}

The sampling overhead of PEC can often be significantly reduced by neglecting the presence of antinoise outside the causal lightcone of an observable \cite{tran2023locality}.
Noise and antinoise outside the light cone do not, by definition, affect the measurement outcomes, but such antinoise, if included in the construction, artificially increases the statistical variance.
The analysis in Ref.~\cite{tran2023locality} employed a binary-valued definition of a causal lightcone: an error is either inside or outside the cone.
Here, we generalize the notion of a causal lightcone to a continuous version we call a ``shaded lightcone'' (Fig.~\ref{fig_slc}a), providing tighter upper bounds on the observable bias resulting from each error channel.
These causal bounds endow PEC with some of the sampling-efficiency of other observable-aware QEM methods, with no loss of generality or rigor. Going further, we enable additional sampling-cost reductions by considering how errors interact with either the state or the observable, classically evolving noise perturbations backwards or forwards in time via the interaction picture.
These classical computations are exponentially expensive, and the realizable benefit of our algorithm may be limited by the number of simulable layers, determined by properties such as the density of non-Clifford gates. 
For noise channels beyond the reach of exact simulation, we provide an additional algorithm yielding a looser, but efficiently computable, bound of the bias. These algorithms result in a more targeted application of PEC providing lower variance with only a controlled, and often negligible, effect on accuracy.

The paper is structured as follows, loosely following the steps of the algorithm summarized in Fig.~\ref{fig_slc}b.
After briefly reviewing PEC (Sec. \ref{sec_PEC}), we describe how the bias resulting from the insertion of a single error channel can be upper bounded by unequal-time commutators (Sec. \ref{sec_commutators}), then adapt this result to the case of multiple error channels (Sec. \ref{sec_interactions}). Computing a subset of these commutators (Sec. \ref{sec_fullevolution}) produces a partial shaded lightcone, which is further extended using computationally efficient speed-limit arguments (Sec. \ref{sec_speedlimits}) similar to, but tighter than, Lieb-Robinson bounds \cite{Lieb1972}.
The resulting shaded lightcone enables optimization of PEC for a circuit given a fixed sampling budget or accuracy tolerance.
We describe one such optimization strategy in \cref{sec_prioritization}.
Finally, in \cref{sec:numerics}, we numerically demonstrate our strategy to mitigate the errors in the time evolution of the transverse-field Ising model in one-dimension and two-dimensions, finding a significant reduction in the PEC sampling overhead for this problem.

\section{Probabilistic Error Cancellation}\label{sec_PEC}

In PEC, one wishes to estimate expectation values of a quantum circuit comprised of a sequence of ideal quantum gates, $\{U_l\}$. For each ideal quantum gate $U_l$, we model its realization $\tilde{U}_l$ on a noisy quantum processor by a composition with a noise channel $\Lambda_l$, such that $\tilde{\mathcal U}_l = \Lambda_l \circ \mathcal U_l$. 
Here, $\mathcal U$ denotes the channel version of a unitary $U$.

PEC requires knowledge of the error rates that constitute each $\Lambda_l$~\cite{Van_den_Berg2023-vw}. If $U_l$ is Clifford, such as CNOT or CZ gates, then randomized ``twirling'' with single-qubit Pauli gates \cite{PhysRevLett.76.722, knill2004faulttolerant} permits modeling the channels as Pauli channels on average.
This gives the decomposition $\Lambda_l(\rho) = \Circ_\sigma \left((1-p_{l,\sigma})\rho + p_{l,\sigma} \sigma \rho \sigma\right)$, where each $\sigma$ is a non-identity Pauli occurring independently with respective probability $p_{l,\sigma}$.
While a general noise channel has exponentially many parameters, a tractable and physically motivated model can be obtained by restricting to sparse models with independent 2-local Pauli errors, which has been sufficient to mitigate noise channels in recent experiments \cite{Van_den_Berg2023-vw, Kim2023}. Accurate noise characterization remains a topic of research, particularly due to confounding effects of state-preparation and measurement (SPAM) error \cite{Chen2023-ua}; here we will assume the noise model has been learned accurately.

With the noise channels characterized, one prepares in PEC many copies of the original circuit, and in each deliberately injects errors throughout the circuit with the same probabilities $p_{l,\sigma}$ at which they occur on the noisy hardware. Each time a local error is inserted, an additional minus sign is associated with that copy of the circuit, and these overall signs are included when computing averages from the measurements. The negation is mathematically equivalent to the injection of errors with \emph{negative} probabilities, and on average this so-called ``antinoise'' exactly cancels the bias in the estimation of any observable. However, the cancellation of positive and negative circuits also shrinks the resulting expectation values by a factor of $\gamma = \prod_{l,\sigma}(1-2p_{l,\sigma})^{-1}$. Multiplying by $\gamma$ recovers unbiased mitigated estimates, but with a statistical variance also increased by $\gamma^2$, and one must increase the number of samples accordingly to recover the expectation values up to a fixed precision.

The PEC sampling cost $\gamma^2$ grows exponentially in the size of the circuit. Notably, all antinoise throughout the circuit contributes uniformly to $\gamma^2$, regardless of how much the corresponding noise channels ultimately impact the measured observable. Ref.~\cite{tran2023locality} noted that neither noise nor antinoise outside the causal lightcone of an observable affects the expectation value, and that the sampling cost can be significantly reduced by neglecting these terms. Below, we describe how a closer inspection of the interactions of the state, errors, and observable permit further reductions in sampling cost.

\section{Bounding the bias from an error}\label{sec_commutators}

Suppose one tries to prepare an ideal state $\rho$ in order to estimate the expectation value of an observable $A$, but a Hermitian error $E$ occurs during the quantum circuit. How much does the error bias the expectation value?

\begin{figure}[t]
    \includegraphics[width=0.85\linewidth]{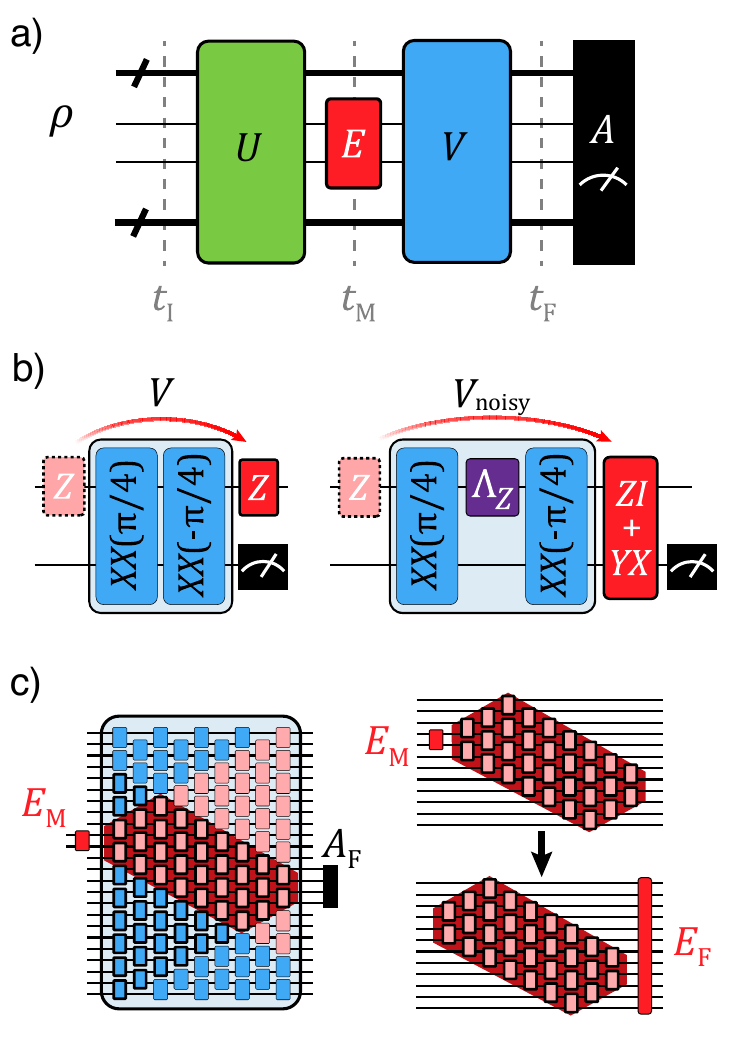}
    \caption{a) An estimate of $\braket{A}$ is biased by an error $E$ that occurs at time $t\subM$. If the error can be numerically exactly evolved forwards through $V$ to time $t\subF$, or backwards through $U$ to time $t\subI$, then the bias can be bounded using the commutator with $A$ or with $\rho$. b) These commutators may be decreased or increased by the presence of other noise, which must be accounted for carefully to obtain an upper bound. In this example, $E$ (red) would commute with $A$ (black) except for the component of $E$ scattered from another noise channel, $\Lambda_Z$. c) The unequal-time commutator depends only on those gates in the intersection of the two operators' future (pink) and past (bold) lightcones. This intersection (dark red) determines the size of the operator that must be computed.}
    \label{fig_setup}
\end{figure}

The answer is given by the unequal-time commutators between $A$, $\rho$, and $E$. 
As in the interaction picture of quantum mechanics, each of these three operators can be evolved forwards or backwards in time through the other operations comprising the circuit, and we will use the subscripts $\textrm{I}$, $\textrm{M}$, $\textrm{F}$ to denote an operator thus evaluated at the start of the circuit (initial), at the time the error occurs (middle), or at the end of the circuit (final), respectively (\cref{fig_setup}a).
Typically, one knows in advance the initial state $\rho\subI = \ket{0}\hspace{-2pt}\bra{0}$, the error when it occurs $E\subM$, and the observable operator at the time of the measurement $A\subF$. Suppose, temporarily, that there are no other noise sources in the circuit, and all operations are unitary. Then the Hermitian error $E$ biases the estimate of $\langle A \rangle$ by
\begin{align}
    \mathrm{Bias}_E(A) &= \mathrm{Tr}(A\subF E\subF \rho\subF E\subF) - \mathrm{Tr}(A\subF \rho\subF) \nonumber \\
    &= \text{Tr}([E\subF, \rho\subF] [E\subF, A\subF])/2.
    \label{bias_schrodinger}
\end{align}
Because the trace is invariant under unitary time evolution of the operators, $(\rho, E, A) \rightarrow (U\rho U^\dagger, UEU^\dagger, UAU^\dagger)$, it can be evaluated given operator values simultaneous at any time $t\in\{\textrm{I}, \textrm{M}, \textrm{F}\}$,
\begin{equation}\label{bias_actual}
    \mathrm{Bias}_E(A) = \mathrm{Tr}([E_t, \rho_t] [E_t, A_t])/2.
\end{equation}
This equation also applies for more general circuits with non-unitary operations provided the evolution of $E\subM$ to $E_t$ is unitary (App. \ref{sec:evolution_condition}), and we will work to ensure this condition in Sec. \ref{sec_interactions}.
Generically, simultaneous values for all three operators are not available; if they were, one would not need a quantum computer to estimate $\braket{A}$. Fortunately, an upper bound can be obtained from only two. By H\"older's inequality,
\begin{equation}
    \abs{\mathrm{Bias}_E(A)} \leq \norm{[E_t, \rho_t]}_n \norm{[E_t, A_t]}_m/2,
    \label{main_bound}
\end{equation}
which holds for Schatten norms satisfying $1/n+1/m=1$. Choosing $(n,m)=(1,\infty)$ prevents the $m$-norm from becoming large, ensuring both $\norm{[E_t, \rho_t]}_1 \leq 2$ and $\norm{[E_t, A_t]}_\infty \leq 2$. This choice is natural since the Schatten 1-norm (nuclear norm) is related to the trace distance, and the Schatten $\infty$-norm (spectral norm) reflects a worst-case choice of input state. Thus if either $E\subI$ or $E\subF$ can be computed from $E\subM$, it can be used to compute an upper bound, $\left\Vert[E\subI, \rho\subI]\right\Vert_1$ or $\left\Vert[E\subF, A\subF]\right\Vert_\infty$, where the unknown norm in Eq.~\eqref{main_bound} has been replaced by the trivial bound of 2.

If we further assume that the rest of the circuit is noiseless, then the unitary invariance of Schatten norms permits evaluating the two commutators in Eq.~\eqref{main_bound} at different times, such that
\begin{equation}\label{bias_bound_unitary}
    \abs{\mathrm{Bias}_E(A)} \leq \norm{[E\subI, \rho\subI]}_1 \norm{[E\subF, A\subF]}_\infty/2.
\end{equation}
This equation can also be applied in special cases such as circuits comprised of only Clifford gates and Pauli channels (\cref{sec:clifford_bound}), or when all errors besides $E$ have been mitigated. In qualifying problems where both $E\subI$ and $E\subF$ can be computed (but not necessarily $\rho\subF$ nor $A\subI$), Eq.~\eqref{bias_bound_unitary} provides a tighter bound than Eq.~\eqref{main_bound}.

More generally, to bound the bias resulting from the insertion of an error channel of the form $\Lambda(\rho) = (1-p)\rho + pE\rho E$ with $E$ Hermitian, the relevant bound for $\mathrm{Bias}_E(A)$ is simply multiplied by the error rate $p$; doing so for Eq.~\eqref{main_bound} gives
\begin{equation}
    \abs{\mathrm{Bias}_\Lambda(A)} \leq p\norm{[E_t, \rho_t]}_1 \norm{[E_t, A_t]}_\infty/2.
    \label{main_channel_bound}
\end{equation}


\section{Accounting for error-error interactions} 
\label{sec_interactions}
Above, we considered evolving a single error, associated with a single error channel, through a circuit to bound the bias introduced by inserting that error. In reality, a noisy quantum circuit contains many such error channels. The total bias, $\mathrm{Bias}(A)$, thus depends on a complex cascade of error-error interactions produced by this arrangement of channels. The presence of one channel can even increase the effect of another (\cref{fig_setup}b). One might accordingly expect the task of obtaining a similar bound on the overall effect of many error channels to be much more difficult. Happily, useful bounds on the total bias can be obtained while sidestepping this complexity entirely.

To bound $\mathrm{Bias}(A)$, we begin with the ideal circuit, and construct the noisy circuit by using Eq.~\eqref{main_channel_bound} to insert error channels one-by-one. Let $\{\Lambda_i(\rho) = (1-p_i)\rho + p_iE_i\rho E_i\}$ be a list of all error channels in the circuit, and $\braket{A}_j$ be the expectation value of the circuit including all error channels $i\leq j$, such that $\braket{A}_0$ is the ideal result and $\braket{A}_N$ is that with all $N$ noise channels included. The total bias is the sum of the incremental biases introduced by each additional channel:
\begin{gather}
\mathrm{Bias}(A) = \braket{A}_N - \braket{A}_0 = \sum_{i=1}^N\braket{A}_{i} - \braket{A}_{i-1} \nonumber, \\
\abs{\mathrm{Bias}(A)} \leq \sum_{j=1}^N\abs{\braket{A}_j - \braket{A}_{j-1}} \label{eq:inductive_bias}
\end{gather}
By Eq.~\eqref{main_channel_bound}, the $j$th term is bounded by
\begin{equation}\label{eq:incremental_bias}
\abs{\braket{A}_j - \braket{A}_{j-1}} \leq p_j\norm{[(E_j)_t, \rho_t]}_1 \norm{[(E_j)_t, A_t]}_\infty/2.
\end{equation}
To use Eq.~\eqref{eq:incremental_bias}, we will need to obtain either $(E_j)\subI$ or $(E_j)\subF$ by evolving $E_j$ through the circuit that includes all the previously inserted channels $\{\Lambda_{i<j}\}$.
However, our derivation of Eq.~\eqref{eq:incremental_bias} via Eq.~\eqref{bias_actual} required that this evolution of $E_j$ must be unitary.
A solution is to time order the list $\{\Lambda_i\}$, with the ordering defined by the requirement that the evolution of each $E_j$ to the end of the circuit containing only $\{\Lambda_{i<j}\}$ is unitary. Bounding $\norm{[(E_j)\subF, \rho\subF]}_1$ by 2, we obtain the total bound
\begin{equation}
\mathrm{Bias}(A) \leq \sum_j p_j \norm{[(E_j)\subF, A\subF]}_\infty,
\end{equation}
which may be computed by evolution of each error $E_j$ forwards through the remainder of the ideal circuit. The opposite time-ordering of $\{\Lambda_i\}$ provides the analogous bound with $\norm{[(E_j)\subI, \rho\subI]}_1$, though this is typically looser. For either choice, the list of commutator norms, or upper-bounds thereof, may be computed in advance without knowledge of the hardware error rates, and then the overall bound trivially completed as the dot product with $\{p_i\}$ once those error rates are available.

A small generalization of this ordering procedure (\cref{sec:partitioned_evolution}) provides an improved bound,
\begin{align}\label{eq:ordered_bound}
\mathrm{Bias}(A) \leq &\sum_{j\leq T} p_j \norm{[(E_j)\subI, \rho\subI]}_1 + \sum_{j>T} p_j \norm{[(E_j)\subF, A\subF]}_\infty
\end{align}
for any non-negative $T \leq N$ partitioning the time-ordered error channels into those evolved backwards and those evolved forwards. For deep circuits where $E_j$ can be classically evolved through relatively few layers, this bound becomes insensitive to the choice of $T$, and produces the same result regardless of whether the $j$th term (Eq.~\eqref{eq:incremental_bias}) was bounded using the general-case (Eq.~\eqref{main_bound}) or special-case (Eq.~\eqref{bias_bound_unitary}) expression. For shallower circuits, both $T$ and the ordering of mutually-commuting error channels may be chosen to minimize $\mathrm{Bias}(A)$. In our implementation we generate this partition using a straightforward greedy algorithm, which, though not optimal, runs efficiently enough to be applied quickly after the noise model is obtained. This approach yields sensible partitions in the examples studied here.

If the entire circuit is composed of Clifford gates, then Pauli errors remain Pauli errors regardless of where they are propagated to in the circuit. Since a later Pauli channel simply dampens the effect of an earlier Pauli error, we can leverage \cref{bias_bound_unitary} instead of selecting a cutoff $T$. See Appendix~\ref{sec:clifford_bound}.

\section{Lightcone shading: Computational methods for bounding the bias} 
\label{sec_computingbounds}

The bounds for all possible errors in a circuit form a shaded lightcone. We define the value of the shaded lightcone at the channel with error $E_j$ to be $\norm{[(E_j)\subI, \rho\subI]}_1$ where $j\leq T$, and $\norm{[(E_j)\subF, A\subF]}_\infty$ where $j>T$, unless the circuit contains only Clifford gates in which case for all $j$ we use the tighter bound $\norm{[(E_j)\subI, \rho\subI]}_1 \norm{[(E_j)\subF, A\subF]}_\infty/2$. When a commutator cannot be computed, we replace it with the tightest available upper bound. The tighter the bounds, the more efficiently one can apply PEC to estimate $\braket{A}$. We now present a combination of classical methods for bounding the bias via \cref{main_bound}. 
Though the complexity of computing the unequal-time commutator norms grows exponentially with circuit depth, the interaction picture permits several helpful optimizations (\cref{sec_fullevolution}).
When this computation is no longer feasible, we use the classically efficient algorithm detailed in \cref{sec_speedlimits} to extend these results deeper into the circuit.

\subsection{Classical evolution of $E$}\label{sec_fullevolution}
Computing $E\subI$ or $E\subF$ by evolving $E\subM$ is possible for errors sufficiently near the beginning or end of the circuit. We computationally represent an arbitrary error $E$ as a sum of Pauli matrices, since they form an operator basis. $E$ often remains small during evolution, either in operator weight (number of qubits with non-identity Paulis) or in the number of nonzero terms in the Pauli basis representation. For example, a Pauli error can be efficiently evolved through a Clifford circuit due to the lack of growth in Pauli space. 
Assuming a 2-local Pauli noise model, each $E\subM$ is a weight-one or weight-two Pauli error, 
which depending on the circuit structure can be numerically evolved through $\sim$10 or more non-Clifford gates
with modest computational resources before the operator becomes too large for a laptop computer.
To obtain $E\subF$, one need only evolve $E\subM$ through the intersection of its future lightcone with the past lightcone of $A\subF$ (\cref{fig_setup}c), which can significantly reduce the necessary operator size.
In principle, a single backwards evolution of $A\subF$ could be reused to compute commutators with many errors $E\subM$, but the lack of any forward lightcone in this error-agnostic, Heisenberg-picture evolution leads to much larger operators.

Besides time evolution, evaluation of the commutator norms can also be computationally limiting. For errors evolved backwards, the nuclear norm $\norm{[E\subI, \ket{0}\hspace{-2pt}\bra{0}]}_1$ can be computed relatively quickly in the Pauli basis (\cref{sec:nuclearnorm}), but for forward evolution, the spectral norm $\norm{[E\subF, A\subF]}_\infty$ is the largest singular value of $[E\subF, A\subF]$, which is more difficult. This step, which we perform in the computational basis using a sparse implementation \cite{qrusty, pyscf} of Davidson's method \cite{Davidson}, limits the depth from which errors can be profitably evolved forwards in the example circuits analyzed here. Details of how we restrict to sufficiently small operators accompany the example in \cref{sec:tfim1d}. Nonetheless, the resulting shaded lightcone can still be extended further into the interior of the circuit by the efficient classical computation described below.

\subsection{Information-theoretic speed limits} \label{sec_speedlimits}

So far we have classically evolved errors forwards to compute or bound $\norm{[E\subF,A\subF]}_\infty$ where computationally feasible. Now we switch to the perspective of evolving $A$ backwards. By the unitary invariance of Schatten norms, we may reinterpret the previous results as bounds on $\norm{[E\subM,A\subM]}_\infty$. Since we know $E\subM$, we can solve for new bounds on the local Pauli components of $A\subM$, providing partial information about $A\subM$ even though we never evolved $A\subF$ backwards. Inspired by the ideas behind the Lieb-Robinson bounds, which upper bound the speed of information propagation in quantum systems, we now describe an algorithm that efficiently evolves this partial information about $A\subM$ to even earlier times in the circuit, allowing us to compute bounds on the bias due to even earlier errors.

Recall that, in our notation, $A\subM = V^\dag A V$ (Fig.~\ref{fig_setup}) is the observable propagated to where the error $E\subM = E$ happens.
If the qubits are embeded on a lattice and $V\approx e^{-iH \tau}$ is a unitary that approximates the time evolution of a geometrically local Hamiltonian $H$ on this lattice, the Lieb-Robinson bound~\cite{Lieb1972} states that
\begin{align} 
     \norm{[E,V^\dag A V]}_\infty \lesssim e^{v_\textrm{LR}\tau - r_{EA}},
\end{align}
where $r_{EA}$ is the spatial distance between the support of $E$ and $A$ and $v_{\textrm{LR}}$ is the Lieb-Robinson velocity.
The Lieb-Robinson bound effectively defines an operator-spreading lightcone $r_{EA} \lesssim v_\textrm{LR}\tau$ outside of which the bias introduced by the error $E$ on $A$ is negligible.
So, in principle, one can readily use the Lieb-Robinson bound and its generalizations to arbitrary connectivity graphs \cite{Chen_2021, Chen_2023} to bound the bias in \cref{main_bound}.
However, the Lieb-Robinson bound is insensitive to the commutativity between the terms of the Hamiltonian, making it very loose in many scenarios. 
In particular, the Lieb-Robinson velocity is nonzero even when the Hamiltonian consists of only mutually commuting terms.

Given an error operator $E$ and a decomposition of $V = V_1\dots V_L$ into $L$ one- and two-qubit gates $V_1,\dots, V_L$, our algorithm introduces ``local bounds'' $w_{\ell,i,\sigma}$ ($\sigma = x,y,z$), which upper bound the $\sigma$ component on qubit $i$ of the operator $E$ propagated through $\ell$ gates. 
Intuitively, these local bounds $w_{\ell,i,\sigma}$ provide an operator-spreading lightcone of $A$ under $V$ similar to the Lieb-Robinson bounds.
However, in contrast to derivations of Lieb-Robinson bounds that use the worst-case bounds to propagate the light cone, our algorithm uses the Pauli transfer matrices of $V_1,\dots,V_L$ to iteratively compute $w_{\ell+1,i,\sigma}$ from $w_{\ell,i,\sigma}$.

We denote by $A^{(\ell)} = V_\ell^\dag\dots V_1^\dag A V_1 \dots V_\ell$ the operator $A$ propagated through the first $\ell$ gates.
For each site $i$, we can always decompose $A^{(\ell)}$ as
\begin{align} 
     A^{(\ell)} = \sum_{\sigma \in I, X, Y, Z} \sigma \otimes A^{(\ell)}_{\sigma,[i]},
\end{align}
where $\sigma$ acts only on site $i$ and $A^{(\ell)}_{\sigma,[i]}$ are operators supported possibly everywhere but on site $i$.
Our algorithm returns local bounds $w_{\ell,i,\sigma}$ such that 
\begin{align} 
     \norm{A^{(\ell)}_{\sigma,[i]}}_\infty \leq w_{\ell,i,\sigma},
\end{align}
for all $\ell,i,\sigma$.
To compute $w_{\ell,i,\sigma}$ iteratively, we use the following lemma:

\begin{lemma}\label{lem:iterative_bound}
    Let $i,j$ be the support of a two-qubit gate $V_\ell$.
    Let $W^{(\ell)} \in \mathbb R^{16}\times \mathbb R^{16}$ be the Pauli transfer matrix of $V_\ell$, i.e.
    \begin{align} 
        V_\ell^\dag \sigma_i \otimes \tau_j V_\ell = \sum_{\sigma',\tau'} W^{(\ell)}_{\sigma\tau,\sigma'\tau'} \sigma'_i \otimes \tau'_j, \label{eq:ptm}
    \end{align}
    where $\sigma,\tau \in {I, X, Y, Z}$.
    We have
    \begin{align} 
        \norm{A^{(\ell)}_{\sigma,[i]}}_\infty
        \leq \sum_{\tau} \sum_{\sigma',\tau'}  \abs{W^{(\ell)}_{\sigma'\tau',\sigma\tau}} \min\{w_{\ell-1,i,\sigma'},w_{\ell-1,j,\tau'}\}.\label{eq:w_ell+1}    
    \end{align}
\end{lemma}

We present a proof of this lemma in \cref{sec:lemmaproof}.
Since the Pauli transfer matrix involves at most two qubits for each gate, the upper bound in \cref{eq:w_ell+1} can be computed efficiently. Choosing $w_{\ell,i,\sigma}$ to be the right-hand side of \cref{eq:w_ell+1}, \Cref{lem:iterative_bound} provides an iterative algorithm to compute the local bounds.
Although we state the lemma for two-qubit gates, it also applies to one-qubit gates by simply adding a fictitious qubit to the system.

Given the Pauli decomposition of the observable $A$, \Cref{lem:iterative_bound} provides the iterative procedure to efficiently compute the local bounds $w_{\ell,i,\sigma}$ as we propagate $A$ through the circuit.
The local bounds in turn provide upper bounds on the commutator $\norm{\comm{E, V^\dag A V}}_\infty$.
For example, if $E = \sigma_i$ is a Pauli matrix supported on only site $i$, we have
\begin{align} 
     \norm{\comm{E,V^\dag A V}}_\infty
     = \norm{\sum_\tau\comm{\sigma_i, \tau_i}\otimes A^{(L)}_{\tau,[i]}}_\infty
     \leq \sum_{\tau \neq \sigma, I} w_{L,i,\tau}.
\end{align}
This bound can be generalized to operators $A$ being arbitrary Pauli strings or linear combinations of Pauli strings using the chain rule for commutators and the triangle inequality.

To understand how the local bounds take into account the gate commutativity when the Lieb-Robinson bound fails to do so, we can consider a toy example where the circuit consists of only Pauli $ZZ$ rotations on nearest-neighbors in a one-dimensional lattice of $n$ qubits:
\begin{align} 
    V = \left[\prod_{i=0}^{n-2} e^{-i Z_i Z_{i+1} \theta }\right]^k, 
\end{align}
where $\theta$ is a constant.
This circuit is the Trotterized time evolution of a 1D Ising model and $k$ plays the role of the number of Trotter steps.
Consider an initial observable $A = X_0$.
While applying the Lieb-Robinson bound to this circuit would result in a lightcone supported on a number of qubits proportional to $\theta k$, the gates in $V$ are mutually commuting and should spread the supported $A$ to only the second qubit.

In contrast, this behavior is well captured by the Pauli transfer matrices $W^{(\ell)}$ defined in \cref{eq:ptm}. In this example, these matrices have the property that
$W^{(\ell)}_{\sigma \tau, \sigma'\tau'} = 0$ if $\sigma, \tau \in \{Z, I\}$ and $\sigma' \neq \sigma$ or $\tau' \neq \tau$.
In other words, $W^{(\ell)}$ never changes the Pauli type of the operator if it initially consists of only $Z$ and $I$.
We start with the initial local bounds $w_{0,i,\sigma}$ which is nonzero only if $\sigma = X$ at $i = 0$ or $\sigma = I$ at $i\neq 0$.
After the first gate $e^{-iZ_0 Z_1 \theta}$, the additional possibly nonzero local bounds are $w_{1,0,Y}$ and $w_{1,1,Z}$.
Using the recursive relation \cref{eq:w_ell+1}, we can find the local bounds after the second gate $e^{-i Z_1 Z_2\theta}$. 
In particular, for qubit 2, we have
\begin{align} 
     w_{2,2,\sigma} = \sum_{\tau} \sum_{\sigma',\tau'}  \abs{W^{(2)}_{\sigma'\tau',\sigma\tau}} \min\{w_{1,2,\sigma'},w_{1,1,\tau'}\}.
\end{align}
Recall that $w_{1,2,\sigma'} = 0$ unless $\sigma' = I$ and, similarly, $w_{1,1,\tau'} = 0$ unless $\tau' \in \{Z, I\}$.
The property of $W^{(\ell)}$ mentioned earlier enforces $\sigma = \sigma' = I$, resulting in  $w_{2,2,I}$ as the only possible nonzero local bound on qubit 2.
It implies that the evolved version of $A$ cannot have nontrivial support on qubit 2 and, by following this recursive relation, any qubits other than 0 and 1.
The local bounds thus recover the correct constant-size lightcone under the circuit $V$.

\section{Allocation of antinoise}\label{sec_prioritization}

We now turn our attention from the construction of the shaded lightcone to the application of it in PEC. Compute time on quantum devices is scarce, limiting an experiment to a fixed number of shots. For PEC, this limit corresponds to a fixed budget of antinoise that can be distributed over the different error sources. Prior calculations show how to bound the impact of a particular error on the bias on the final observable. How can we use this information to allocate our antinoise budget to achieve the tightest rigorous bound on the final bias on the expectation value?

Suppose all noise channels are guaranteed to take the form of Pauli-Lindblad noise due to twirling. An error in this model after a gate $U_l$ at noise rate $\lambdals$ corresponds to a Pauli error $\sigma$ occurring with probability $p(\lambdals) = (1-e^{-2\lambdals})/2$. At the site of the error channel $\Lambda_l =  \Circ_\sigma  e^{\lambdals \mathcal{L}_{\sigma}}$, with Lindbladian $\mathcal{L}_{\sigma}(\rho) := \sigma\rho \sigma - \rho$, we may insert a non-positive antinoise channel $e^{-\lambdals^* \mathcal{L}_{\sigma}}$, with an antinoise rate $\lambdals^* \leq \lambdals$. This antinoise reduces the effective noise rate to $\lambdals- \lambdals^*$ at a cost of increasing the variance of the PEC estimator by a factor $e^{4\lambdals^*}$. Selecting $\lambdals^* > \lambdals$ not only increases the variance more than necessary but also rapidly introduces additional bias.

After applying antinoise, we have a collection of Pauli-Lindblad error channels $e^{(\lambdals-\lambdals^*) \mathcal{L}_{l,\sigma}}$ throughout the circuit. 
For each $l, \sigma$, we have computed a bound $c_{l,\sigma} \geq \abs{\text{Bias}(A)}$ on the bias on the final observable $A$ induced by an error $\sigma$ after $U_l$ on its own---that is, $c_{l,\sigma}$ is the shaded lightcone. Since each error occurs with probability $p(\lambdals-\lambdals^*)$, the contribution to the bias of each error is $p(\lambdals-\lambdals^*) c_{l,\sigma}$. By the triangle equality, the total bias is upper bounded by
\begin{align}
\sum_{l,\sigma} p(\lambdals-\lambdals^*) c_{l,\sigma}
&=\sum_{l,\sigma} \frac{1 -e^{-2(\lambdals - \lambdals^*)} }{2} c_{l,\sigma}.\label{eq:total_bias}
\end{align}
The limited antinoise budget imposes that the allocation of the $\lambdals^*$ must satisfy 
\begin{align} 
     \sum_{l,\sigma} \lambdals^* \leq C, \quad \text{and}\ 0 \leq \lambdals^* \leq \lambdals, \label{eq:constraints}
\end{align}
where $C$ is a constant.
Therefore, finding the optimal antinoise distribution reduces to minimizing the total bias in \cref{eq:total_bias}, subject to the constraints in \cref{eq:constraints}.
Minimizing \cref{eq:total_bias} is equivalent to maximizing the expression
\begin{align}
\sum_{l,\sigma} e^{2\lambdals^*} \underbrace{c_{l,\sigma} e^{-2\lambdals}}_{\equiv \alpha_{l,\sigma}} =  \sum_P e^{2\lambdals^*} \cdot \alpha_{l,\sigma}\label{eq:convex_opt}
\end{align}
where we view $\alpha_{l,\sigma}$ as the \emph{priority} of the noise source $P_i$ after $U_l$.

Note that the priority of a noise source depends on both the value of the shaded lightcone and its noise rate. It may appear counter-intuitive that the higher the noise rate $\lambdals$ at which the error occurs, the \emph{lower} its priority. However, there is a simple interpretation of this phenomenon by viewing $\alpha_{l,\sigma}$ as a measure of the quality of an investment of a small amount of antinoise. With no antinoise, the bias is proportional to $1 - e^{-2\lambdals}$. The investment quality is the slope of this function, which is $2e^{-2\lambdals}$. As $\lambdals$ increases, the investment quality becomes exponentially close to $0$. Therefore, errors with large $\lambdals$ may be considered ``too far gone'' and not worth mitigating.  

 To maximize \cref{eq:convex_opt}, we observe that the expression is increasing in all $\lambdals^*$, 
 and that the problem is convex. Thus the solution occurs at a vertex of the polytope formed by the hypervolumes of $\lambdals^* \leq \lambdals$ and $\sum_{l,\sigma} \lambdals^* \leq C$. Hence, all $\lambdals^*$ except at most one satisfy either $\lambdals^* = 0$ or $\lambdals^*= \lambdals$. The allocation is readily obtained by sorting the noise sources in decreasing order of $\alpha_{l,\sigma}$ and fully mitigating as many high-priority sources as is within budget. Once no more noise sources can be mitigated fully, one more can be mitigated partially.

While the shaded lightcone exhibits a continuous measure of error bias, selecting a noise model and imposing an antinoise budget produces an (almost) binary-valued antinoise allocation. In this sense, a shaded lightcone and noise model may be viewed together as a collection of discrete antinoise allocations that can be interpolated between depending on the available sampling budget.

\section{Numerical examples\label{sec:numerics}}

\begin{figure*}[t]
    \includegraphics[width=0.8\textwidth]{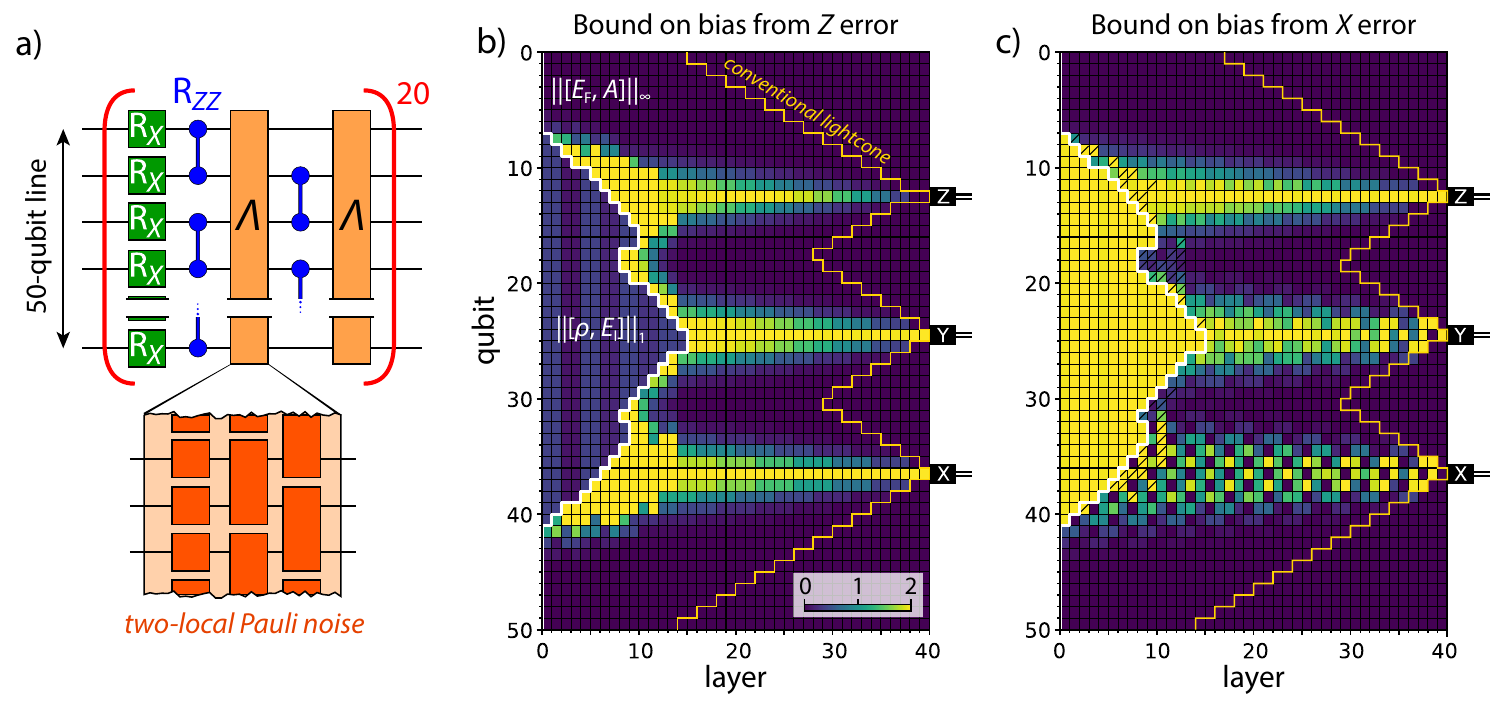}
    \caption{Shaded lightcones for the noisy one-dimensional transverse-field Ising model Trotter circuit in (a), with 50 qubits, 40 layers of two-qubit $R_{ZZ}$ gates, gate angles $\theta_X = \pi/16$ and $\theta_{ZZ} = -\pi/2$, and observable $A = X_{12}Y_{24}Z_{36}$. Each pixel bounds the bias contribution of an individual (b) $Z$ or (c) $X$ error somewhere in the circuit. Dark pixels indicate errors that would inflict little or no bias if neglected in PEC. A conventional lightcone based on commutation checks would fill all points left of the yellow boundary with the trivial bound of 2 (Fig.~\ref{fig_slc}(a)). The white boundary separates bounds on the effect of the error on the initial state or on the observable, corresponding to the time-ordering partition $T$ in Eq.~\eqref{eq:ordered_bound}. Slashes indicate commutator weights exceeding $N_{\mathrm{max}}$, where a looser triangle-inequality bound was computed instead of $\norm{[E\subF, A\subF]}_\infty$. The complete shaded lightcone includes 12 such plots (\cref{sec:tfim1d_supp}), collectively bounding all possible two-local Pauli errors. Moderate thresholds $B_{\mathrm{max}}=5 \cdot 10^5$, $N_{\mathrm{max}} = 20$ were chosen for this pedagogical figure to enable computation on a laptop. This set of computations completed in $\sim10$ hours on a laptop; trivial parallelization over some of the 23,640 error channels in the circuit could reduce this time (by a smaller factor), at a cost of more memory.
    }
    \label{fig:tfim1d}
\end{figure*}

To show how the above methods fit together in a software implementation, along with the expected sampling-cost benefits, we numerically demonstrate example applications of the lightcone-shading technique. Following recent benchmarks of quantum and classical methods of estimating expectation values \cite{Kim2023, anand2023classical}, we target the Trotter-evolution circuit of the transverse-field Ising model Hamiltonian on a heavy-hex lattice. In two dimensions, this model is non-integrable, so cannot in general be efficiently simulated on a classical computer~\cite{Mondaini_2016}. As a pedagogical warm up, we first analyze the one-dimensional transverse-field Ising model, walking through features of the circuit setup and shaded-lightcone analysis in this simpler system, before turning to the full problem on heavy-hex topology.

\subsection{Transverse-field Ising model in 1D}\label{sec:tfim1d}
The transverse-field Ising model Hamiltonian,
\begin{equation}
    H = -J\sum_{i<j} Z_iZ_j + h\sum_i X_i,
\end{equation}
describes a spin lattice with nearest-neighbor interaction strength $J$ and a global transverse field $h$. In the first-order Trotter circuit describing the time evolution of this system (\cref{fig:tfim1d}a), each step consists of a layer of $R_X$ gates with angle $\theta_X = 2h\Delta_t$ composed with $R_{ZZ}$ gates with angle $\theta_{ZZ} = -2J\Delta_t$; we fix $\theta_{ZZ} = -\pi/2$ to match \cite{Kim2023}.

\begin{figure}[t]
    \includegraphics[width=0.95\linewidth]{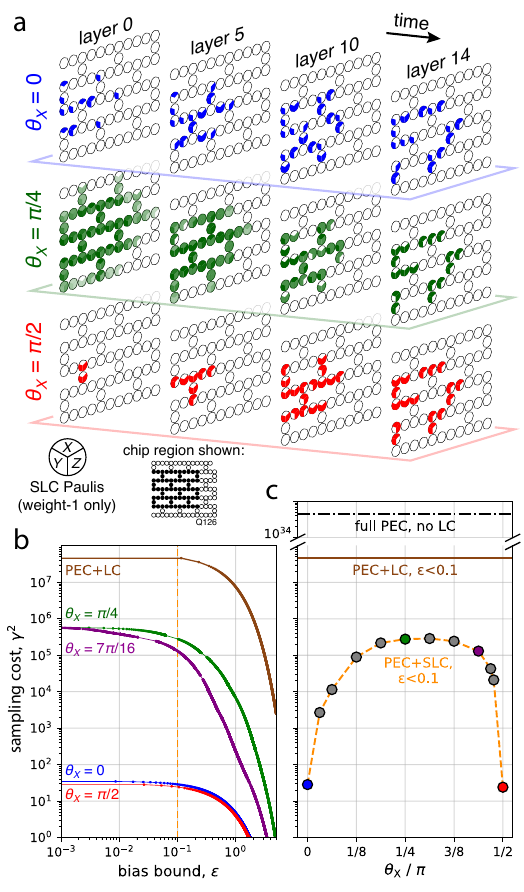}
    \caption{Sampling costs of the heavy-hex transverse-field Ising model circuit. \textbf{a)} Each circular sector represents a weight-1 component (legend) of the shaded lightcone for the five Trotter step circuit with $\theta_X = 0, \pi/4, \pi/2$, at times immediately following the indicated layer of CNOTs. Colors indicate values from 0 (white) to 2 (solid). \textbf{b)} Moving right to left on this plot corresponds to using PEC to mitigate more errors using the shaded lightcone for each choice of $\theta_X$ (colors) or using a traditional lightcone (brown), trading off bias (bounded by horizontal axis value) for variance (proportional to vertical axis). \textbf{c)} Sampling costs needed to bound the bias below 0.1, approximately at the intersections with the dashed orange line in (b). For this example, the lightcone shading provides a $>$150-times improvement in sampling cost for all $\theta_X$.}
    \label{fig:utility5}
\end{figure}

We first analyze the circuit in \cref{fig:tfim1d}a: a line of 50-qubits undergoing 20 Trotter steps with $\theta_X = \pi/16$, followed by measurement of the weight-3 observable $X_{36}Y_{24}Z_{12}$. We model the noise as occurring immediately after each layer of gates. Assuming Pauli twirling of the layers \cite{Bennett_twirling, knill_twirling} and that errors are generated locally on individual qubits or nearest-neighbor pairs, the noise model for a single layer reduces to a composition of $3\cdot 50=150$ single-qubit Pauli channels and $9\cdot 49=441$ two-qubit Pauli channels. Each local channel can be specified by the error Pauli $E\subM$, the layer index $\ell$, the spatial index $i$, and the error probability $p$. For mathematical convenience, $p$ may be replaced by a Lindblad error rate $\lambda$, where $p = (1 - e^{-2\lambda})/2 \leq 1/2$. We numerically represent and evolve each error using the \texttt{quantum-info} module of the Qiskit software package \cite{qiskit2024}.

The lightcone shading computation consists of forward evolution to obtain or bound $\norm{[E\subF, A\subF]}_\infty$ for each possible Pauli error $E\subM$ throughout the circuit; extension of these bounds to earlier times (smaller $t\subM$) by speed-limit arguments; and backward evolution to obtain $\norm{[E\subI, \rho\subI]}_1$. Once the noise-model $\{p_i\}$ is known, a fast, greedy optimization of the ordering and partitioning of error channels in Eq.~\eqref{eq:ordered_bound} merges the forward- and backward-bounds into a single set of bounds. (For Clifford circuits, one simply multiplies the two commutator norms per Eq.~\eqref{bias_bound_unitary}, without needing to know $\{p_i\}$). The computed bounds on the bias due to $Z$ errors and $X$ errors, respectively, are displayed in \cref{fig:tfim1d}b,c. Analogous plots for the ten other two-local errors appear in supplementary \cref{sec:tfim1d_supp}.

To regulate the exponential difficulty of lightcone shading, computations are ended when operators grow too large in either Pauli space or real space. For a given type of error $E\subM$, forward evolution is performed iteratively for sites $(\ell,i)$ within the naive causal lightcone of the observable, starting with errors occurring near the end of the circuit $\ell=39$ where forward evolution is trivial, and restarting one layer earlier (layer $\ell-1$). Typically, this process eventually produces an operator with a size $B$ exceeding a user-specified maximum size $B_{\mathrm{max}}$. We define $B$ as the number of boolean entries in the array representation of the operator, approximately twice the number of qubits times the number of terms in the Pauli basis. When this happens, the computation is terminated for that combination of $E\subM$ and $i$. Terminal values of $\ell$ appear in \cref{sec:tfim1d_supp}.

Evaluation of $\norm{[E\subF, A\subF]}_\infty$ can be much more difficult than the time evolution itself, as the evaluation is performed in the computational basis, losing much of the benefit of Pauli-basis sparsity, making the difficulty more sensitive to $N$. Accordingly, whenever the weight $N$ of the time-evolved commutator exceeds a second threshold $N_{\mathrm{max}}$ (slashes in \cref{fig:tfim1d}c), $\norm{[E\subF, A\subF]}_\infty$ is replaced with the one-norm of coefficients of the commutator in the Pauli basis, which is easy to compute. Though looser, this bound remains useful in some regimes.

The three bright peaks centered about the three measurements illustrate how the errors take time to spread across qubits, or reversely how the observable gradually grows as it evolves backwards through the circuit. Notably, the shaded lightcone spreads more slowly than the naive lightcone, the latter determined by the quantum-circuit topology and by which gates commute, which grows at a rate of two qubits per Trotter step due to the commutativity of consecutive $R_{ZZ}$ gates \cite{Kim2023}. This narrowing of the lightcone reflects one way in which lightcone shading produces tighter bounds compared to standard lightcone tracing.

In \cref{fig:tfim1d}b, the shaded lightcone dims just before the $Z$ measurement, as these errors remain near-$Z$ after forward evolution and thus nearly commute with the observable. A similar effect appears in \cref{fig:tfim1d}c just before the $X$ measurement, though the effect is obscured by the nontrivial action of $R_{ZZ}$ gates on $X$.

As described in \cref{sec_speedlimits}, the gate angles set speed limits on the flow of information through real- and Pauli-space, and these enable an extension of the previously-computed shaded lightcone at negligible computational cost. This makes a pronounced improvement in \Cref{fig:tfim1d}b in the difficult regions where $B > B_{\mathrm{max}}$ (App. \ref{sec:tfim1d_supp}).

Next, the time-reversed version of the forward-evolution algorithm is performed to compute $\norm{[E\subI, \rho\subI]}_1$, beginning with errors $\ell=0$ and iteratively restarting at larger $\ell$ until encountering $B>B_{\mathrm{max}}$ (App. \ref{sec:tfim1d_supp}). This norm is readily computed in the Pauli basis (App. \ref{sec:nuclearnorm}), so no threshold on $N$ is needed.  The benefit of this computation is typically small, as the component of an error that commutes with $\ket{0}\bra{0}$ after backwards evolution tends to drop off quickly with $\ell$, but can be significant for errors occurring sufficiently early in circuits with near-Clifford gates (\cref{fig:tfim1d}b).

Finally, the two sets of bounds are merged into a single set of bounds. This is performed by choosing a suitable space-time boundary $T$ to minimize Eq.~\eqref{eq:ordered_bound}. For this example, we assume the simple case of uniform noise rates $\{\lambda_i\}$, and our greedy optimization determines the white boundary in the figures. The boundary taper ensures that $\rho$-commutator bounds are never used within the forward lightcone of channels where $A$-commutator bounds are used (nor the reverse), allowing use of Eq.~\eqref{eq:ordered_bound}.

\subsection{Transverse-field Ising model in heavy-hex}\label{sec:tfim2d}
To illustrate the application of lightcone shading to PEC, we consider the 127-qubit, depth-15 circuit from \cite{Kim2023, anand2023classical} consisting of five Trotter-steps of the transverse-field Ising model Hamiltonian on a two-dimensional heavy-hex lattice followed by measurement of the weight-17 observable $X_{37,41,52,56,57,58,62,79} Y_{75} Z_{38,40,42,63,72,80,90,91}$.  \Cref{fig:utility5}a shows the components of the shaded lightcone that bound effects of weight-one errors, computed for this circuit for three values of $\theta_x$. For experimental compatibility, $R_{ZZ}(-\pi/2)$ gates are compiled using a CNOT as in \cite{Kim2023}, modeling the Pauli noise as immediately following each CNOT layer, which effectively rotates the noise layer by single-qubit Cliffords compared to the definition in \cref{fig:tfim1d}a. Each shaded lightcone, including the weight-two components (not shown), was computed on a laptop in roughly an hour each with no parallelization of the main loop over error channels, with computational thresholds $B_\mathrm{max}=10^6$ and $N_\mathrm{max}=20$. At earlier times in the non-Clifford $\theta_X = \pi/4$ circuit, the shaded lightcone spreads outward from the measurements, reflecting the spatial growth of the backwards-evolved observable $A\subM$ (not computed). Here $A\subI$ would extend slightly beyond the layer-0 shaded lightcone due to the action of the initial two-qubit gate layer. In contrast, at $\theta_X = 0$, the commutativity of $R_{ZZ}$ gates restricts the shaded lightcone to the vicinity of the measured qubits at all times. And at $\theta_X = \pi/2$, the shaded lightcone \emph{shrinks} at earlier times, because in this example $A\subI$ is by construction the weight-1 $Z$ on qubit 58. The shaded lightcones visually indicate how circuits of equal size may be more or less difficult to mitigate: a larger and denser shaded lightcone defines a larger region where consequential errors might occur, leading to a larger mitigation sampling cost.

To estimate realistic sampling costs, we use a Pauli noise model learned on quantum hardware during the relevant experiment in \cite{Kim2023}, and suppose an accuracy tolerance $\varepsilon < 0.1$, noting that the actual bias \cref{bias_actual} may be smaller. With these inputs, the procedure in \cref{sec_prioritization} produces the sampling costs in \cref{fig:utility5}(b,c). For comparison, performing full PEC with no consideration of a lightcone has an intractable sampling cost $\gamma^2 = 4\cdot 10^{34}$. A conventional lightcone, combined with our prioritization scheme to obtain $\varepsilon<0.1$, dramatically reduces this cost to $5\cdot 10^7$. Unlike a conventional lightcone, the shaded lightcone takes into account the action of each gate, and the resulting sampling cost thus varies with gate angle $\theta_X$. For all values of $\theta_X$, lightcone shading reduces the sampling cost of obtaining $\varepsilon < 0.1$ to less than $3\cdot10^5$, more than a factor of 150 below the conventional lightcone result, enabling the application of PEC to this circuit in less than a day given current job execution speeds on IBM systems. Significant improvements in execution speed should be possible via further optimization of classical software or the use of a field programmable gate array (FPGA) for circuit compilation \cite{fruitwala2024}, enabling application to yet more difficult problems.

\section{Conclusion and Outlook}\label{sec_conclusion}

Lightcone shading enables quantum error mitigation of larger problems while maintaining the rigorous accuracy bounds of PEC, and opens many promising avenues of research. As errors tend to decay as they evolve through subsequent error channels, accounting for this decay, even partially, may tighten the bias bounds significantly or yield more efficient mitigation strategies. A more efficient computation of the operator norm, particularly if it can be performed without leaving the sparse Pauli basis, might relieve that classical bottleneck. More generally, new advances in classical methods for simulating quantum circuits may in turn be used for lightcone shading, potentially enabling larger error-mitigated quantum computations. One promising optimization based on truncating Pauli terms with small coefficients while retaining exact bias bounds \cite{CPT2023} may facilitate deeper computations. It may also be possible to use shaded lightcones to productively eliminate noise-channels from mitigation methods besides PEC, such as simplifying the network in tensor-network error mitigation \cite{filippov2023scalable}. Multiple compatible observables may be estimated from a single dataset by repeatedly analyzing the dataset and choosing different subsets of error channels to treat as antinoise each time \cite{upanddownpatentapp}, at the cost of doubling the Lindblad rates of the unmitigated, lower-priority channels. Theoretically connecting shaded lightcone computations to the quantum error correction literature of decoders, which also track the effect of propagated quantum errors on specific measurements, may prove fruitful in unifying aspects of error mitigation and error correction research programs, particularly with an eye towards layering both approaches. Finally, by providing a window between microscopic operator dynamics and the difficulty of performing error mitigation, lightcone shading stands to provide enabling insights for problem selection in the ongoing pursuit of near-term quantum advantage.

\begin{acknowledgments}
    We thank Jeffrey Cohn for bringing to our attention the importance of error-error interactions with the example in Fig. \ref{fig_setup}b. We thank Ewout van den Berg and Luke Govia for their detailed comments on a draft of the manuscript. We thank Bryce Fuller, Christopher Wood, Samantha Barron, Mario Motta, Will Kirby, Kunal Sharma, and Abhinav Kandala for helpful conversations, technical assistance, and programmatic support.
\end{acknowledgments}

\textbf{Competing interests:} Elements of this
work are included in a patent application filed by the International
Business Machines Corporation with the US Patent and
Trademark Office.

\appendix

\section{Derivation of Equation \eqref{bias_actual}}\label{sec:evolution_condition}
We consider the general circuit structure in Fig. \ref{fig_setup}a, but replace $U$ and $V$ with general channels $\Lambda_1$ and $\Lambda_2$, which are not necessarily unitary.

We know in advance the initial state $\rho\subI = \rho$, the Hermitian error when it occurs $E\subM = E$, and the observable operator at the time of the measurement $A\subF = A$. The biased expectation can always be found, in principle, by evolving $\rho$ forwards through the circuit including $E$ (Schr{\"o}dinger picture),
\begin{equation}
    \mathrm{Bias}_E(A) = \mathrm{Tr}\Big(A\Lambda_2(E\Lambda_1(\rho)E)\Big) - \braket{A}_0,
\end{equation}
where $\braket{A}_0$ is the expectation without the error $E$, or by evolving $A$ backwards through the circuit including $E$ (Heisenberg picture),
\begin{equation}
    \mathrm{Bias}_E(A) = \mathrm{Tr}\Big(\rho\Lambda_1^\dag(E\Lambda_2^\dag(A)E)\Big) - \braket{A}_0.
\end{equation}
Decomposing the two channels in terms of Kraus operators $\{K_k\}$ and $\{L_l\}$ gives, for either picture,
\begin{align}\label{eq:appKraus}
    \mathrm{Bias}_E(A) = 
    &\mathrm{Tr}\Big(A\sum_{k,l} L_l E K_k \rho K_k^\dag E L_l^\dag \Big) -\braket{A}_0,
\end{align}
When does that expression equal the following expression from Eq.~\eqref{bias_schrodinger}?
\begin{equation}
    \mathrm{Tr}(A\subF E\subF \rho\subF E\subF) - \braket{A}_0
\end{equation}
Plugging the definitions,
\begin{align}
    \rho\subF &= \Lambda_2(\Lambda_1(\rho)) = \sum_{k,l} L_l K_k \rho K_k^\dag L_l^\dag, \\
    E\subF &= \Lambda_2(E) = \sum_{l} L_{l} E L_{l}^\dag,
\end{align}
 into Eq.~\eqref{eq:appKraus} gives
\begin{equation}
    \mathrm{Tr}\Big(A \sum_{k l l' l''} L_{l'} E L_{l'}^\dag L_l K_k \rho K_k^\dag L_l^\dag L_{l''} E L_{l''}^\dag \Big) - \braket{A}_0.
\end{equation}
The desired cancellations $L_{l'}^\dag L_l = 1$ and $L_l^\dag L_{l''} = 1$ occur if $\Lambda_2$ consists of only a single Kraus operator, i.e. that $\Lambda_2$ is unitary. This justifies Eq.~\eqref{bias_schrodinger}, and thus also Eq.~\eqref{bias_actual} for $t=\mathrm{F}$, provided $\Lambda_2$ is unitary. A slight modification of the above argument justifies the case $t=\mathrm{I}$ provided $\Lambda_1$ is unitary, and the case $t=\mathrm{M}$ for arbitrary $\Lambda_1, \Lambda_2$.

The core issue is that in our classical computation of $\rho_t, E_t, A_t$ we time-evolve each operator independently, which can miss correlations imprinted on these operators by the fact that the same random noise acts on each -- not just identically random noise, but identical random noise. For example, if $E_t$ was classically computed by evolving $E_M$ through $\Lambda$, then one of either $\rho_t$ or $A_t$ was classically computed using an identical, but independent, copy of $\Lambda$. In contrast, in the quantum computation, the very same \emph{instance} of $\Lambda$ acts on both $E_t$ and the other operator. Thus our classical computation includes terms where $\Lambda$ applies, e.g. an $X$ error during the evolution of $E$ but no error during the evolution of $\rho$, which does not describe reality: if the channel yields an $X$ error on one run of the circuit, then both $E$ and $\rho$ are acted on by that same $X$ error. However, if $E_t$ is the result of unitary evolution of $E_M$, then no two of $\rho_t, E_t, A_t$ depend on any common noisy channel $\Lambda$, so each may be classically computed independently without missing effects of noise correlations.

\section{Bounding the bias of a Clifford circuit with Pauli noise}\label{sec:clifford_bound}
Consider a circuit composed of a sequence of Clifford gates $\{C_i\}$ and Pauli channels $\{\Lambda_i\}$, with initial state $\rho$.
For brevity we will write the Clifford gates as channels, $\mathcal{C}_i$, and write channel composition as multiplication, $\mathcal{C}\mathcal{D} = \mathcal{C}\circ\mathcal{D}$.
The expectation value of Pauli $A$ can be written explicitly as
\begin{equation}
    \braket{A} = \mathrm{Tr}\big(A (\Circ_i \mathcal{C}_i \Lambda_i)[\rho]\big).
\end{equation}
We wish to bound the bias due to the insertion of another Pauli channel $\mathcal{E}$; one may choose $\mathcal{E}[\rho] = (1-p)\rho + p E \rho E$ to match the analysis in the main text. For definiteness, say $\mathcal{E}$ occurs just after $C_{i_0}$. Then we wish to bound the magnitude of
\begin{align}
    \mathrm{Bias}(A) =\ &\mathrm{Tr}\big(A (\Circ_{i>i_0} \mathcal{C}_i \Lambda_i)\mathcal{E} (\Circ_{i\leq i_0} \mathcal{C}_i \Lambda_i)[\rho]\big) \nonumber \\
    - &\mathrm{Tr}\big(A (\Circ_{i>i_0} \mathcal{C}_i \Lambda_i)(\Circ_{i\leq i_0} \mathcal{C}_i  \Lambda_i)[\rho]\big).
\end{align}
A Pauli channel evolved through a Clifford gate remains a Pauli channel, and Pauli channels commute with one another. By thus evolving all Pauli channels to the end of the circuit, we can write the expectation value in terms of a new set of Pauli channels $\{\Lambda_i'\}$,
\begin{align}
    \mathrm{Bias}(A) =\ &\mathrm{Tr}\big(A (\Circ_{i} \Lambda_i')(\Circ_{i>i_0} \mathcal{C}_i)  \mathcal{E}  (\Circ_{i\leq i_0} \mathcal{C}_i)[\rho]\big) \nonumber \\
    - &\mathrm{Tr}\big(A (\Circ_{i} \Lambda_i')(\Circ_{i} \mathcal{C}_i) [\rho]\big).
\end{align}
To allow the channels to act on $A$, we rewrite the expectation values in the Heisenberg picture,
\begin{align}
    \mathrm{Bias}(A) =\ &\mathrm{Tr}\big(\rho (\Circ_{i\leq i_0} \mathcal{C}^\dagger_i)  \mathcal{E}  (\Circ_{i > i_0} \mathcal{C}_i^\dagger)  (\Circ_{i} \Lambda_i')[A]\big) \nonumber \\
    - &\mathrm{Tr}\big(\rho (\Circ_{i} \mathcal{C}_i^\dagger)  (\Circ_{i} \Lambda_i') [A]\big),
\end{align}
noting that $\Lambda = \Lambda^\dagger$ for a Pauli channel. The Pauli observable $A$ is an eigenvector of the composite Pauli channel with overall Pauli fidelity $f \leq 1$,
\begin{align}
    \mathrm{Bias}(A) =\ f\Big(&\mathrm{Tr}\big(\rho (\Circ_{i\leq i_0} \mathcal{C}^\dagger_i)  \mathcal{E}  (\Circ_{i > i_0} \mathcal{C}_i^\dagger) [A]\big) \nonumber \\
    - &\mathrm{Tr}\big(\rho (\Circ_{i} \mathcal{C}_i^\dagger) [A]\big)\Big).
\end{align}
The expression in parentheses is precisely the bias due to the insertion of error channel $\mathcal{E}$ into the otherwise-noiseless version of the circuit, and thus for the choice $\mathcal{E}[\rho] = E\rho E$ is bounded in magnitude by Eq.~\eqref{bias_bound_unitary} if we define $E\subI, E\subF$ as the results of evolving $E$ through only the noiseless circuit operations. 

Of course, for such circuits the exact bias (not to mention the ideal, noiseless expectation value) can also be classically computed efficiently, so the bounds here may have more theoretical than practical value.

\section{Tightening the bound by evolving early errors backwards and late errors forwards}\label{sec:partitioned_evolution}

As in \cref{sec_interactions}, we will bound $\mathrm{Bias}(A)$ by starting from the ideal circuit, and constructing the noisy circuit by inserting error channels one-by-one, using using Eq.~\eqref{main_channel_bound} and the triangle inequality to update the bound on the total bias at each step. We distinguish between the time ordering $t \in \{1, ..., N\}$ at which errors occur in the circuit, and the order $i \in \{1, ..., N\}$ in which we insert errors into the ideal circuit to construct the noisy circuit. We insert all errors with $t\leq T$ in reverse time-order, then insert those with $t>T$ in forward time-order, so the list of errors in insertion order is 
\begin{equation}
\{ \Lambda_{i=1}^{t=T}, \Lambda_{i=2}^{t=T-1}, ..., \Lambda_{i=T}^{t=1}, \Lambda_{i=T+1}^{t=T+1}, \Lambda_{i=T+2}^{t=T+2}, ..., \Lambda_{i=N}^{t=N}\}.
\end{equation}

As before, we let $\braket{A}_j$ be the expectation value of the circuit including all error channels $i\leq j$, such that $\braket{A}_0$ is the ideal result and $\braket{A}_N$ is that with all $N$ noise channels included. The total bias is the sum of the incremental biases introduced by each additional channel, such that applying the triangle inequality gives the upper bound:
\begin{gather}
\abs{\mathrm{Bias}(A)} \leq \sum_{j\leq T}\abs{\braket{A}_j - \braket{A}_{j-1}} + \sum_{j>T}\abs{\braket{A}_j - \braket{A}_{j-1}}. 
\end{gather}
By Eq.~\eqref{main_channel_bound}, and recalling that each norm is individually less than or equal to 2,
\begin{equation}\label{eq:partitioned_bias}
\abs{\mathrm{Bias}(A)} \leq \sum_{j\leq T} p_j\norm{[(E_j)\subI, \rho\subI]}_1 + \sum_{j>T} p_j\norm{[(E_j)\subF, A\subF]}_\infty,
\end{equation}
where in each term $E_j$ must be evolved to either the start or end of the respective circuit containing the error channels $\{\Lambda_{i<j}\}$. The specific time-ordering of the list $\{\Lambda_i\}$ ensures that this evolution never involves evolving $E_j$ through any error channel $\Lambda_i$, and thus all evolutions can equivalently be performed with respect to the ideal circuit, with no further consideration of the time-ordering. This preserves the important feature that each commutator is independent of the error rates $\{p_j\}$, i.e. that the bound depends only linearly on $\{p_j\}$.

For a sufficiently deep circuit, we can define $t_\mathrm{early}$ and $t_\mathrm{late}$, such that $\norm{[(E_j)\subI, \rho\subI]}_1$ is accessible via classical computation only for $j$ where $t < t_\mathrm{early}$, and likewise $\norm{[(E_j)\subF, A\subF]}_\infty$ is accessible only for $j$ where $t > t_\mathrm{late}$. For $E_j$ not satisfying these conditions, the best we can do is to replace the associated commutator norm with the looser, trivial bound of 2. For a circuit sufficiently deep that $t_\mathrm{early} < t_\mathrm{late}$, then for any choice of $T$ between $t_\mathrm{early}$ and $t_\mathrm{late}$, Eq.~\eqref{eq:partitioned_bias} reduces to: 
\begin{align}
\abs{\mathrm{Bias}(A)} \leq &\sum_{\{j|t < t_\mathrm{early}\}} p_j\norm{[(E_j)\subI, \rho\subI]}_1 \nonumber \\
&+ \sum_{\{j|t_\mathrm{early}<t<t_\mathrm{late}\}} 2p_j \nonumber \\
&+ \sum_{\{j|t>t_\mathrm{late}\}} p_j \norm{[(E_j)\subF, A\subF]}_\infty.
\end{align}
This result is insensitive to the choice of partition time $T$, and also equivalent to using the triangle inequality to combine the result of the special-case bound Eq.~\eqref{bias_bound_unitary} for each $E_j$ when that computation is subject to the same computational constraints $t_\mathrm{early}$ and $t_\mathrm{late}$.

\section{Computation of the nuclear norm $\norm{[E\subI, \ket{0}\hspace{-2pt}\bra{0}]}_1$ in the Pauli basis}\label{sec:nuclearnorm}

Here we describe a classical algorithm to compute the commutator norm $\norm{\comm{E\subI,\ket{0}\hspace{-2pt}\bra{0}}}_1$ given a Pauli decomposition of $E\subI$.
We make use of the symplectic representation of an $n$-qubit Pauli operator, $\sigma_{x,z} = (-i)^{x\cdot z}Z^{z}X^{x}$, where $x,z$ are length-$n$ bitstrings. After evolving $E\subM$ backwards to the beginning of the circuit, one has the Pauli-basis representation $E\subI = \sum_{x,z}c_{x,z}\sigma_{x,z}$ and wishes to compute the nuclear norm of the commutator with the initial state $\rho\subI = \ket{0}\hspace{-2pt}\bra{0}$. Discard all Pauli terms where $x=0$ since they commute with $\rho\subI$ and call the remaining sum $E'$. The desired commutator is $C = [E', \ket{0}\hspace{-2pt}\bra{0}] = \ket{\psi}\hspace{-2pt}\bra{0} - \ket{0}\hspace{-2pt}\bra{\psi}$, where $\ket{\psi} = E'\ket{0}$ is orthogonal to $\ket{0}$. 

Define the normalized state $\ket{\overline{\psi}}=\ket{\psi}/\sqrt{s}$ ; after some algebra, one finds the normalization factor:
\begin{equation}
s = \braket{\psi|\psi} = \sum_{x\neq0} \left| \sum_z c_{x,z} i^{z\cdot x}\right|^2,
\end{equation}
which can be computed by first sorting the list of terms by $x$, then computing the inner sum for each section of the list with constant $x$.

The nuclear norm can be written $\norm{C}_1 = \mathrm{Tr}(\sqrt{C^\dag C})$. By the above, $C^\dag C = s(\ket{0}\hspace{-2pt}\bra{0}-\ket{\overline{\psi}}\hspace{-2pt}\bra{\overline{\psi}})$, which is a diagonal matrix with two nonzero elements, both equal to $s$. Thus we have for the nuclear norm,
\begin{equation}
\norm{C}_1 = 2\sqrt{s} \leq 2,
\end{equation}
and similarly for the Frobenius and spectral norms, 
$\norm{C}_2 = \sqrt{\mathrm{Tr}(C^\dag C)} = \sqrt{2s}$ and $\norm{C}_\infty = s$, respectively.

\section{Proof of \cref{lem:iterative_bound}} \label{sec:lemmaproof}
In this section, we present a proof of \cref{lem:iterative_bound} in the main text.

\begin{proof}
Expanding $A^{(\ell-1)} = \sum_{\sigma,\tau} \sigma_i\otimes \tau_j\otimes A^{(\ell-1)}_{\sigma\tau,[i,j]}$ in the Pauli basis on sites $i,j$, where $A^{(\ell)}_{\sigma\tau,[i,j]}$ are some operators supported possibly everywhere except for $i,j$, and using the definition of $W^{(\ell)}$, we have
\begin{align} 
    &V_\ell^\dag A^{(\ell-1)} V_\ell = \sum_{\sigma,\tau,\sigma',\tau'} W^{(\ell)}_{\sigma'\tau',\sigma\tau} \sigma_i \otimes \tau_j \otimes A^{(\ell-1)}_{\sigma'\tau',[i,j]}\nonumber\\
    &= \sum_{\sigma} \sigma_i \otimes \bigg(\underbrace{\sum_{\tau, \sigma',\tau'} W^{(\ell)}_{\sigma'\tau',\sigma\tau}  \tau_j \otimes A^{(\ell-1)}_{\sigma'\tau',[i,j]}}_{= A^{(\ell)}_{\sigma,[i]}}\bigg).
\end{align}
Using the triangle inequality, we have
\begin{align} 
    \norm{A^{(\ell)}_{\sigma,[i]}}
    \leq  \sum_{\tau, \sigma',\tau'} \abs{ W^{(\ell)}_{\sigma'\tau',\sigma\tau}} \norm{A^{(\ell-1)}_{\sigma'\tau',[i,j]}}.\label{eq:inside_lem1_0}
\end{align}
To relate the right-hand side by $w_{\ell-1,\sigma',i}$ and $w_{\ell-1,\tau',j}$, we note that
\begin{align} 
     \norm{A^{(\ell-1)}_{\sigma'\tau',[i,j]}} &\leq \norm{\sum_{\tau'}\tau'_j\otimes A^{(\ell-1)}_{\sigma'\tau',[i,j]}}
     = \norm{A^{(\ell-1)}_{\sigma',[i]}} \nonumber\\
     &\leq w_{\ell-1,\sigma',i},\label{eq:inside_lem1_1}
\end{align}
where we have used the definitions of $E^{(\ell)}_{\sigma',[i]}$ and $w_{\ell,\sigma',i}$.
Similarly, we have 
\begin{align} 
     \norm{A^{(\ell-1)}_{\sigma'\tau',[i,j]}} &\leq \norm{\sum_{\sigma'}\sigma'_i\otimes A^{(\ell-1)}_{\sigma'\tau',[i,j]}} \leq w_{\ell-1,\tau',j}.\label{eq:inside_lem1_2}
\end{align}
Combining \cref{eq:inside_lem1_0,eq:inside_lem1_1,eq:inside_lem1_2}, we arrive at \cref{lem:iterative_bound}.
\end{proof}

\section{Shaded lightcone for 1D transverse-field Ising model}\label{sec:tfim1d_supp}

\begin{figure*}
    \includegraphics[width=0.6\textwidth]{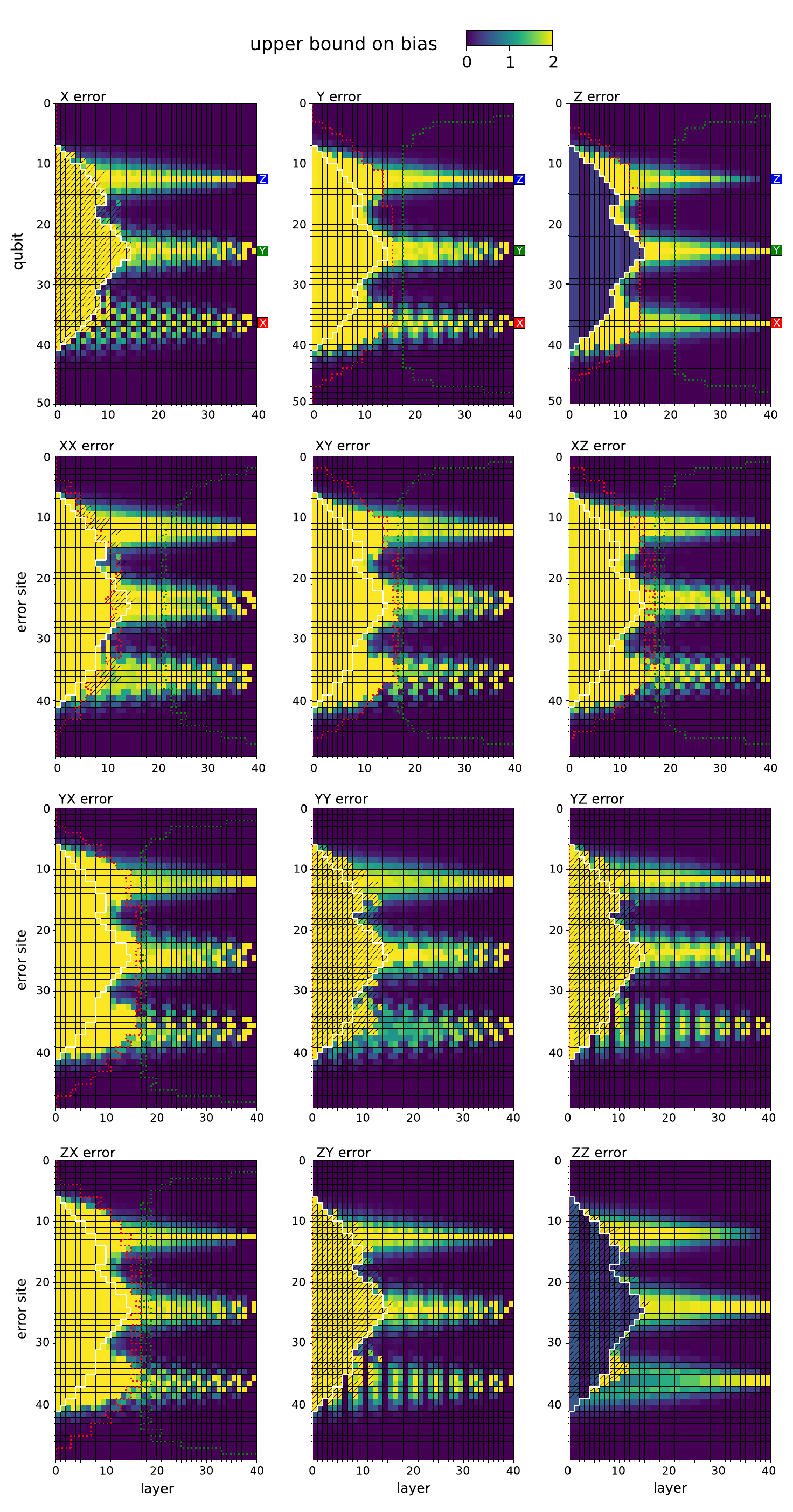}
    
    \caption{All 12 components of the shaded lightcone for the problem discussed in \cref{sec:tfim1d}. As in \cref{fig:tfim1d}, error channels have been ordered with respect to the summations in Eq.~\eqref{eq:ordered_bound} such that channels to the left (right) of the white boundary have $j\leq T$ ($j > T$). The red (green) boundary indicates the earliest (latest) error channel where $E\subF$ ($E\subI$) was computed. Some of these boundaries lie along the very edge of a plot, as in the plot for $X$-error channels (top left). Regions right of the white boundary, but left of the red boundary, were computed exclusively using the speed-limit bound of Sec. \ref{sec_speedlimits}.}
    \label{fig:tfim1d_supp}
\end{figure*}

\Cref{fig:tfim1d_supp} in this section shows all 12 components (3 single- and 9 two-qubit terms) of the shaded lightcone considered in \cref{fig:tfim1d}. The same threshold values $B_\mathrm{max} = 5 \cdot 10^5$ and $N_\mathrm{max} = 20$ are used here.

\clearpage

\bibliography{bib.bib}

\end{document}